\documentclass{PoS}

\usepackage{epsfig}
\usepackage{amsmath,amsfonts,amssymb}
\usepackage{cite}

\newcommand{\ptlep}{p_t^\mathrm{lep}}
\newcommand{\ptmax}{p_t^{j,\mathrm{max}}}
\newcommand{\ptbmax}{p_t^{b,\mathrm{max}}}
\newcommand{\ptmiss}{p_t\!\!\!\!\!\!\!\! \not \,\,\,\,\,\,}
\newcommand{\mthad}{m_T^\mathrm{had}}
\newcommand{\mtlep}{m_T^\mathrm{lep}}

\newcommand{\el}{E_\ell}
\newcommand{\emax}{E_\ell^\text{max}}
\newcommand{\emin}{E_\ell^\text{min}}
\newcommand{\fp}{f_+}
\newcommand{\fm}{f_-}
\newcommand{\fz}{f_0}

\title{New signals in pair production of heavy $Q=2/3$ singlets at LHC}

\ShortTitle{New signals in pair production of heavy $Q=2/3$ singlets at LHC}

\author{\speaker{J. A. Aguilar-Saavedra}\\
        Departamento de F\'{\i}sica Te\'orica y del Cosmos and CAFPE,
	Universidad de Granada, \\ E-18071 Granada, Spain \\
        E-mail: \email{aguilarj@ugr.es}}

\abstract{New quark singlets $T$ can be produced in pairs at LHC through
standard QCD interactions, with a large cross section for masses of several
hundreds
of GeV. Their charged current decays $T \bar T \to W^+ b \, W^- \bar b$, with
semileptonic decay of the $W$ pair,
give a final state $\ell \nu bb jj$, as in top pair production. 
The mixed decay modes
$T \bar T \to W^+ b \, H \bar t,H t \,W^- \bar b \to W^+ b W^- \bar b H$,
$T \bar T \to W^+ b \, Z \bar t,Z t \,W^- \bar b \to W^+ b W^- \bar b Z$,
with $H,Z \to jj$, yield two more jets, $\ell \nu bb jjjj$, or the same
state in the latter case for $Z \to \nu \bar \nu$.
We extend previous work examining in detail the extra $H$ and $Z$ contributions
and stressing one of the salient features of the
$T \bar T \to W^+ b \, W^- \bar b$ signal: the presence of a very energetic
charged lepton. We finally compare with
the discovery potential in $\ell \nu bbbb jj$ final states, with four $b$ tags.}

\FullConference{International Workshop on Top Quark Physics\\
                 January 12-15, 2006\\
                 Coimbra, Portugal}

\begin{document}

\section{Introduction}

Fermion singlets under $\mathrm{SU}(2)_L$ are not present in the minimal
Standard Model (SM); however, their existence is experimentally allowed,
provided their mixing with the standard fermions is of order $O(0.1)$ or
smaller. By far, the best known example in which singlet fermions are
postulated is the
seesaw mechanism, where neutrino singlets are introduced to
explain the smallness of neutrino masses.
Quark singlets appear in some SM extensions like those
with extra dimensions \cite{jose}, in grand unified theories
\cite{barger,frampton}
and Little Higgs models \cite{lhiggs}. In analogy with the neutrino case,
the existence of 
one $Q=2/3$ singlet $T$ (or more) could provide a clue to understand the
largeness of the top mass \cite{paconpb,gui}.

The possible existence of a $Q=2/3$ singlet $T$ with a mass of several hundreds
of GeV would have very interesting phenomenological consequences.
Its mixing with the top and up or charm quarks would generate flavour-changing
neutral couplings at the tree level, large enough to be detectable at the Large
Hadron Collider (LHC) \cite{prl,largo}. In low energy processes, its
contribution could lead to significant departures from SM predictions for
several observables, like rare
kaon decays, $D^0-\bar D^0$ mixing and CP asymmetries in $B$ oscillations
\cite{largo,gustavo,gustavo2,quico}, compatible
with present experimental measurements. But the most interesting ``indirect
effect'' of a quark singlet is Higgs boson discovery
\cite{pacohiggs,pacoprod}: in the presence of a heavy singlet, the luminosity
required for Higgs discovery at LHC is dramatically reduced. For instance, for
$m_T = 500$ GeV, the luminosity needed for discovery of a 115 GeV Higgs is only
6 fb$^{-1}$ \cite{higgs}, ten times smaller than within the minimal SM.

Besides their indirect effects, $Q=2/3$ singlets can be directly observed at
LHC. They can be produced in pairs by QCD interactions
\cite{frampton,pacoprod}, or in association with a light jet, through its mixing
with the bottom quark $V_{Tb}$ \cite{tj,roberto}. Both processes have their own
advantages and weaknesses. Pair production has a model-independent cross section
determined by QCD interactions, which however decreases quickly with $m_T$, due
to phase space suppression. Single $T$ production cross section decreases more
slowly with the heavy quark mass, but on the other hand it is proportional to
$|V_{Tb}|^2$, and gets suppressed for small mixings. Branching ratios for $T
\bar T$ final states involve two heavy quark decays and are thus smaller than
for $Tj$ production. But they also have more fermions in the final state, and
then their backgrounds are less important.
A heavy quark $T$ can decay via charged
currents $T \to W^+ b$, neutral currents $T \to Z t$ and, if the Higgs boson is
light enough, $T \to H t$ as well. The decay widths are approximately
$\mathrm{Br}(T \to W^+ b) = 0.5$,
$\mathrm{Br}(T \to Z t) = 0.25$, $\mathrm{Br}(T \to H t) = 0.25$ (these figures
are exact for $m_T \gg M_W,M_Z,M_H$).
We neglect decays to first and second generation quarks, involving
mixing angles $V_{Ts}$, etc. which are bound to be much smaller than
$V_{Tb}$ \cite{largo}.
The leading decay $T \to W^+ b \to \ell^+ \nu b$ produces a very energetic
charged lepton, which can be used to separate the $T$ signals from their
backgrounds.
The decay mode $T \to Zt \to \ell^+ \ell^- \ell^{'+} \nu b$,
$\ell,\ell' = e,\mu$, gives a
cleaner final state than $T \to W^+ b$, but with a branching ratio 10 times
smaller. The remaining channel $T \to H t \to b \bar b \ell^+ \nu b$ yields 3
$b$ quarks, which can be tagged to reduce backgrounds.
Additionally, it constitutes an important source of Higgs bosons
for moderate $m_T$ values, for which the $T \bar T$ production cross section
is large.

Single $T$ production at LHC has been studied in detail, in the
three decay modes of the heavy quark \cite{azuelos,costanzo}. The best
results have been found for the $T \to W^+ b$ channel. For pair production,
the modes $T \bar T \to W^+ b \, W^- \bar b$ and
$T \bar T \to W^+ b \, H \bar t,H t \,W^- \bar b$ have been
analysed \cite{plb,higgs}. 
Decays to $W^+ b W^- \bar b$, with subsequent semileptonic decay of the $W$
pair, give an experimental signature of one charged lepton, two $b$
jets and at least two additional jets. Decays to 
$W^+ b \, H \bar t,H t \,W^- \bar b$ include two more jets in the final state
from the
Higgs decay, which mostly correspond to $b \bar b$ pairs. The same
occurs in $T \bar T \to W^+ b \, Z \bar t,Z t \,W^- \bar b$
with hadronic $Z$ decay, but with a much smaller fraction of $b \bar b$ pairs.
In the latter mode, the $Z$ boson can also decay invisibly, giving a further
$\ell \nu bbjj$
contribution. In the following we will carefully analyse the additional $H$ and
$Z$ signals, extending what was presented in Ref.~\cite{plb}, and show 
how they enhance the sensitivity in the $W^+ b W^- \bar b$
channel.

\section{Signal and background simulation}
\label{sec:2}

The main backgrounds for the $W^+ b W^- \bar b$ signal and additional
contributions
\begin{eqnarray}
gg,qq & \to & T \bar T \to W^+ b \, W^- \bar b \to \ell^+ \nu b \, \bar q q'
\bar b \;, \quad \quad \ell=e,\mu \,, \nonumber \\
gg,qq & \to & T \bar T \to W^+ b \, H \bar t , H t \, W^- \bar b 
   \to W^+ b \, W^- \bar b \, H \to \ell^+ \nu b \, \bar q q' \bar b 
   \; b \bar b / c \bar c \,, \nonumber \\
gg,qq & \to & T \bar T \to W^+ b \, Z \bar t , Z t \, W^- \bar b 
   \to W^+ b \, W^- \bar b \, Z \to \ell^+ \nu b \, \bar q q' \bar b 
   \; q'' \bar q'' / \nu \bar \nu \,,
\label{ec:sig}
\end{eqnarray}
are given by $t \bar t$, $W b \bar b jj$, $Z b \bar b jj$ and $t \bar b j$
production.
The charge conjugate processes are understood to be summed in all cases.
In our evaluations we do not consider final states with $\tau$ leptons
(which can decay leptonically $\tau \to e \nu_\tau \bar \nu_e$,
$\tau \to \mu \nu_\tau \bar \nu_\mu$) which are estimated to increase
cross sections 3\% or less \cite{higgs}. $Wjjjj$ and $Zjjjj$ production seem to
be negligible after the
requirement of two $b$ tags, which suppresses their cross sections by a factor
$\sim 10^{-4}$. The signal
and the $t \bar t$, $t \bar b j$ backgrounds are evaluated with
our own Monte Carlo generators, including
all finite width and spin effects. We calculate the matrix elements using {\tt
HELAS} \cite{helas} with running coupling constants evaluated at the scale of
the heavy quark $T$ or $t$. $W b \bar b jj$  and $Z b \bar b jj$ are calculated
with {\tt ALPGEN} \cite{alpgen}. The bottom quark mass $m_b=4.8$ GeV is kept
in all cases, and we take $M_H = 115$ GeV. We use structure functions CTEQ5L
\cite{cteq}, with $Q^2 = \hat s$  for $T \bar T$, $t \bar t$ and $t \bar b j$,
and $Q^2 = M_{W,Z}^2 + p_{T_{W,Z}}^2$ for $W b \bar b jj$, $Z b \bar b jj$, 
being $\sqrt{\hat s}$ the partonic centre of mass energy, and $p_{T_{W,Z}}$ the
transverse momentum of the gauge boson.
Events are produced without kinematical cuts at the
generator level in the case of $T \bar T$, $t \bar t$, while for the other
processes we set some loose cuts. For $t \bar b j$
we only require pseudorapidities $|\eta| \leq 3$ for $b$, $j$. For
$W b \bar b jj$ we set
$p_t \geq 15$ GeV and $|\eta| \leq 3$ for the charged lepton, the $b$
quarks and the jets, and lego-plot separations $\Delta R_{jj},\Delta R_{bj},
\Delta R_{bb} \geq 0.4$, $\Delta R_{\ell b},\Delta R_{\ell j} \geq 0.2$. For
$Z b \bar b jj$ we require $p_t \geq 15$ GeV, $|\eta| \leq 3$ for $b$ quarks
and jets, $|\eta| \leq 10$ for the charged leptons, and
$\Delta R_{jj},\Delta R_{bj},\Delta R_{bb} \geq 0.4$.

The events are passed through {\tt PYTHIA 6.228} \cite{pythia} as external
processes to perform hadronisation and include initial and final state radiation
(ISR, FSR). We use the standard {\tt PYTHIA} settings
except for $b$ fragmentation, in which we use the Peterson parameterisation with
$\epsilon_b=-0.0035$ \cite{epsb}. A fast detector simulation {\tt ATLFAST 2.60} 
\cite{atlfast}, with standard settings, is used for the modelling of the ATLAS
detector. We reconstruct jets using a standard cone algorithm with
$\Delta R \equiv \sqrt{(\Delta \eta)^2 + (\Delta \phi)^2} = 0.4$, where $\eta$
is the pseudorapidity and $\phi$ the azimuthal angle. 
The package {\tt ATLFASTB} is used to recalibrate jet energies
and perform $b$ tagging, for which we select efficiencies of 60\%, 50\% for the
low and high luminosity LHC phases, respectively. Efficiencies for charged
lepton identification and triggering are not included.
The simulated events are required to fulfill these two criteria: (a) the
presence of one (and only one) isolated charged lepton, which must have
transverse momentum $p_t \geq 20$ GeV and  $|\eta| \leq 2.5$; (b)
at least four jets with $p_t \geq 20$ GeV, $|\eta| \leq 2.5$, with exactly two
$b$ tags. The presence of the high-$p_t$ charged lepton provides a trigger for
the events.
The cross sections times efficiency of the five processes after these
pre-selection cuts are collected in Table~\ref{tab:1}, using a 60\% $b$
tagging rate. We observe that the additional $H$ and $Z$ contributions to
the signal actually have a larger cross section than the signal itself. For
example, in the first case of $m_T = 500$ GeV we have
$\mathrm{Br} (T \to W^+ b) = 0.503$,
$\mathrm{Br} (T \to H t) = 0.331$, $\mathrm{Br} (T \to Z t) = 0.166$. The
$W^+ b W^- \bar b$ mode has a branching ratio of 0.253 (without including $W$
decays), while the mixed
modes have branching ratios of 0.333 and 0.167, respectively, which have to be
multiplied by extra factors $\mathrm{Br}(H \to jj) = 0.96$ and
$\mathrm{Br}(Z \to jj,\nu \bar \nu) = 0.90$, respectively. 

\begin{table}[htb]
\begin{center}
\begin{tabular}{cl}
Process & $\sigma \times \mathrm{eff}$ \\
\hline
$T \bar T$ (500) & 37.3 fb +  46.5 fb ($H$) + 19.8 fb ($Z$) \\
$T \bar T$ (1000) & 0.618 fb + 0.638 fb ($H$) + 0.481 fb ($Z$) \\
$t \bar t$ & 18.8 pb \\
$W b \bar b jj$ & 1.23 pb \\
$Z b \bar b jj$ & 246 fb \\
$t \bar b j$ & 710 fb
\end{tabular}
\caption{Cross sections of the $T \bar T$ signals (with $m_T = 500,1000$ GeV)
and backgrounds after pre-selection cuts.}
\label{tab:1}
\end{center}
\end{table}

The $T \bar T$ signal can be discovered by the presence of peaks in the
invariant mass distributions corresponding to the two decaying quarks. In order
to reconstruct their momenta we first identify the two jets $j_1$,
$j_2$ from the $W$ decaying hadronically. The first one $j_1$ is chosen to be
the highest $p_t$ non-$b$ jet, and the second one $j_2$ as the non-$b$ jet
having with $j_1$ an invariant mass closest to $M_W$. The missing transverse
momentum is assigned to the undetected neutrino, and its longitudinal momentum
and energy are found requiring that the invariant mass of the charged lepton
and the neutrino is the $W$ mass, $(p_\ell+p_\nu)^2 = M_W^2$. This equation
yields two possible solutions. In addition, there are two different pairings of
the two $b$ jets to the $W$ bosons decaying hadronically and leptonically,
giving four possibilities for the reconstruction of the heavy quark momenta.
We select the one giving closest invariant masses $\mthad$, $\mtlep$ for the
quarks decaying hadronically and semileptonically. Obviously, this
reconstruction procedure does not require the knowledge of the heavy quark mass.

\section{Results for $\boldsymbol{m_T = 500}$ GeV}

\begin{figure}[p]
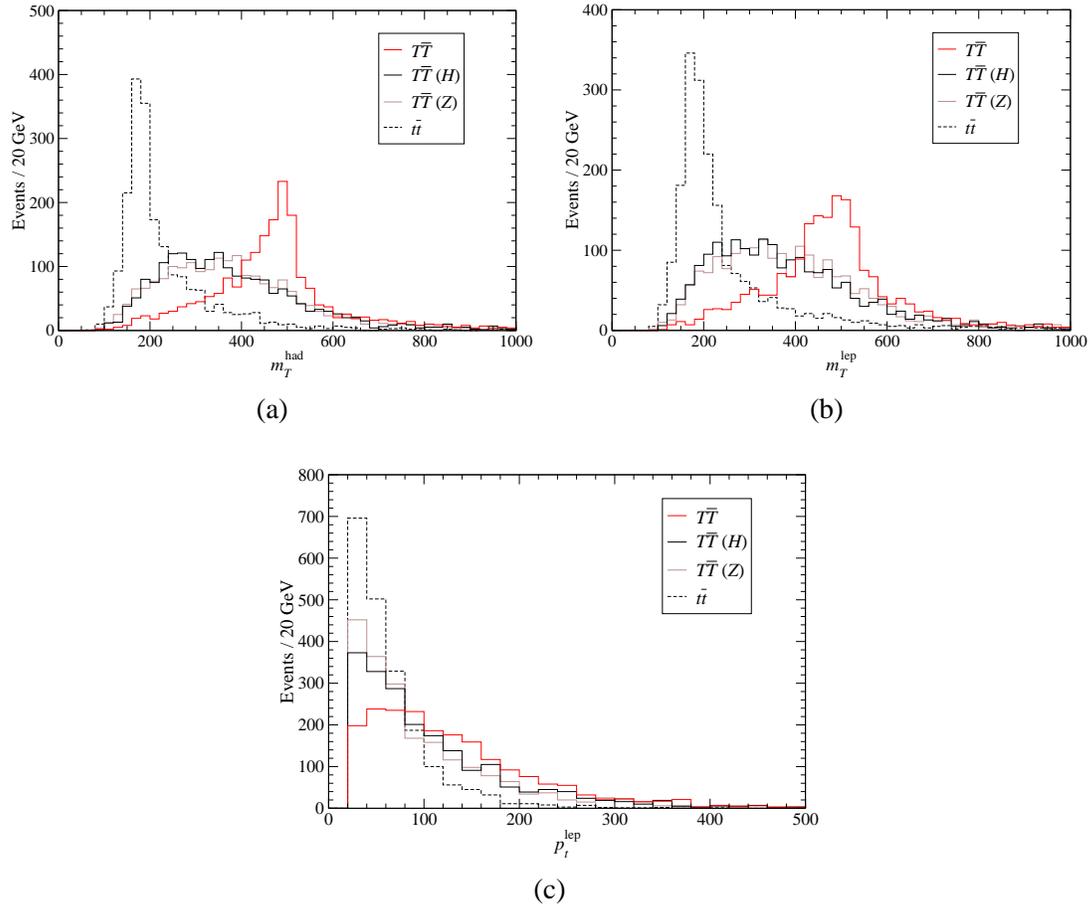

\begin{center}
\begin{tabular}{cc}
\epsfig{file=Figs/mT-had-500.eps,height=5.0cm,clip=} &
\epsfig{file=Figs/mT-lep-500.eps,height=5.0cm,clip=} \\
(a) & (b) \\[0.5cm]
\multicolumn{2}{c}{\epsfig{file=Figs/PTlep-500.eps,height=5.2cm,clip=}} \\
\multicolumn{2}{c}{(c)}
\end{tabular}
\caption{Reconstructed masses of the heavy quarks decaying hadronically (a) and
semileptonically (b), and transverse momentum of the charged lepton (c),
after pre-selection cuts.}
\label{fig:dist500}
\end{center}
\end{figure}

We first present our results for a heavy quark mass of 500 GeV. The
reconstructed masses of the $T$ quarks decaying hadronically and
semileptonically are plotted in Fig.~\ref{fig:dist500} (a) and (b),
respectively, for the $W^+ b W^- b$ signal, the $H$ and $Z$ contributions and
the $t \bar t$ background. All histograms are normalised to
2000 events. The mass reconstruction works reasonably well for $T
\bar T,t \bar t \to W^+ b W^- \bar b$ decays, and can be used to separate them.
For the $H$ and $Z$ signals the distributions are much broader, therefore
eliminating the $t \bar t$ background also suppresses a large fraction of these
signals.
The same happens with the charged lepton transverse momentum $\ptlep$, shown in
Fig.~\ref{fig:dist500} (c). The distribution for the $W^+ b W^- \bar b$ signal
has a long tail, not only due to the higher energy available but also
because the fraction of $W$ bosons with negative helicity produced in $T$ decays
is smaller than for the top quark, and then the charged lepton from $W$ decay is
produced more towards the $W$ flight direction (see section \ref{sec:5}). 
The momentum of the charged lepton can be used to suppress $t \bar t$
production, but unfortunately requiring high $\ptlep$ also reduces
the $H$ and $Z$ contributions, in which the charged 
lepton comes from $t \to W b$ decays half of the time.
Our selection cuts to suppress backgrounds are
\begin{align}
& \ptmax \geq 250 ~\mathrm{GeV} \,, \quad
\ptbmax \geq 150 ~\mathrm{GeV} \,, \nonumber \\
& \ptlep \geq 50 ~\mathrm{GeV} \,, \quad
50 ~\mathrm{GeV} \leq \ptmiss \leq 600 ~\mathrm{GeV} \,, \nonumber \\
& H_t \geq 1000  ~\mathrm{GeV} \,,
\label{ec:cut500}
\end{align}
where $\ptmax$ and $\ptbmax$ are the transverse momenta of the fastest jet and
fastest $b$ jet, respectively; $\ptmiss$ is the missing transverse momentum and
$H_t$ the total transverse energy.  The requirements on these additional
variables do not significantly affect the $H$ and $Z$ contributions.
With these selection cuts in mind, for our high-statistics evaluations we
require at the generator level the presence of a
charged lepton with $p_t \geq 30$ GeV,
a jet with $p_t \geq 200$ GeV and, for $W b \bar b jj$ and $Z b \bar b jj$,
one $b$ quark with $p_t \geq 100$ GeV. This last cut does not bias the sample
because in these two processes the two non-$b$ jets mostly originate from light
quarks and gluons, for which the $b$ mistag probability is very low. Thus, the
$b$-tagged jets correspond to the $b$ quarks most of the time.
For the backgrounds we simulate the number of events corresponding to 10
fb$^{-1}$ of luminosity, and for the signals we simulate 100 fb$^{-1}$
and divide the result by 10, to reduce statistical fluctuations.
Signal and background cross sections are scaled by $K$ factors which
approximately take into account higher order processes with extra jet radiation
(see Ref.~\cite{plb} for details).

\begin{figure}[p]
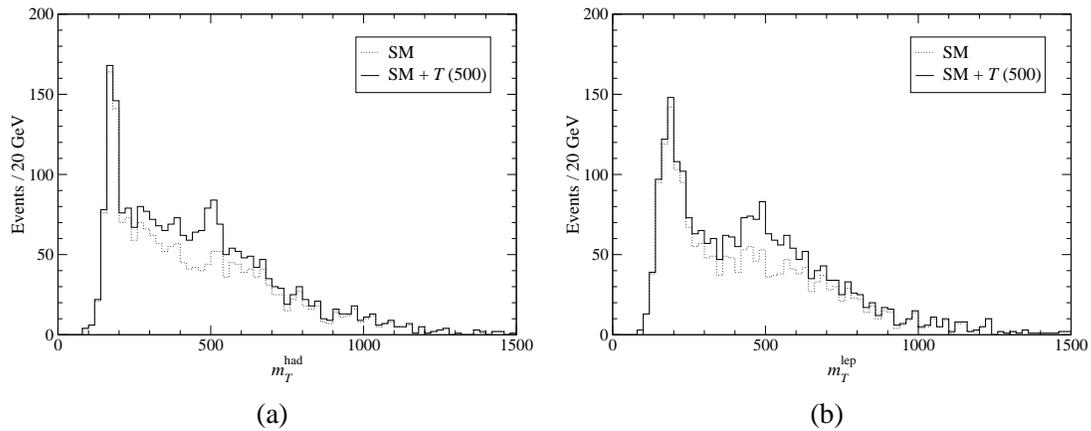

\begin{center}
\begin{tabular}{cc}
\epsfig{file=Figs/mT500-had.eps,height=5.0cm,clip=} &
\epsfig{file=Figs/mT500-lep.eps,height=5.0cm,clip=} \\
(a) & (b)
\end{tabular}
\caption{Reconstructed masses of the heavy quarks decaying hadronically (a) and
semileptonically (b), after selection cuts. The
dashed lines correspond to the SM predictions, while the full lines represent
the SM plus a new 500 GeV quark.}
\label{fig:mT500}
\end{center}
\end{figure}

The mass distributions after selection cuts of
the SM background and background plus signal are shown in Fig.~\ref{fig:mT500}.
The number of events $N_\mathrm{cut}$ corresponding to each process can be read in
Table~\ref{tab:500}. The second column displays the numbers of events in the
peak region, defined as
\begin{equation}
300 ~\mathrm{GeV} \leq \mthad \leq 660 ~\mathrm{GeV} \;, \quad
300 ~\mathrm{GeV} \leq \mtlep \leq 660 ~\mathrm{GeV} \,.  
\label{ec:peak500}
\end{equation}
In this region, the statistical significance of the signals is $\mathcal{S}_0
\equiv S/\sqrt B = 10.94$ or, if we include a 5\% background normalisation
uncertainty, $\mathcal{S}_5 \equiv S/\sqrt{B+(0.05 B)^2} = 7.69$. Discovery of
the new quark would be possible with 2.1 fb$^{-1}$ of luminosity (2.8 fb$^{-1}$
if we include the background uncertainty).

\begin{table}[htb]
\begin{center}
\begin{tabular}{ccc}
Process & $N_\mathrm{cut}$ & $N_\mathrm{peak}$ \\
\hline
$T \bar T$ (500) & 201.7 & 135.5 \\
$T \bar T$ (500, $H$) & 139.4 & 59.5 \\
$T \bar T$ (500, $Z$) & 58.5 & 26.5 \\
$t \bar t$ & 1609 & 300 \\
$W b \bar b jj$ & 287 & 81 \\
$Z b \bar b jj$ & 39 & 13 \\
$t \bar b j$ & 70 & 16 \\
\end{tabular}
\caption{Number of events $N_\mathrm{cut}$ passing the selection criteria and
number of events $N_\mathrm{peak}$ passing the
selection cuts which are in the peak regions.}
\label{tab:500}
\end{center}
\end{table}

\section{Results for $\boldsymbol{m_T = 1}$ TeV}

\begin{figure}[p]
\begin{center}
\begin{tabular}{cc}
\epsfig{file=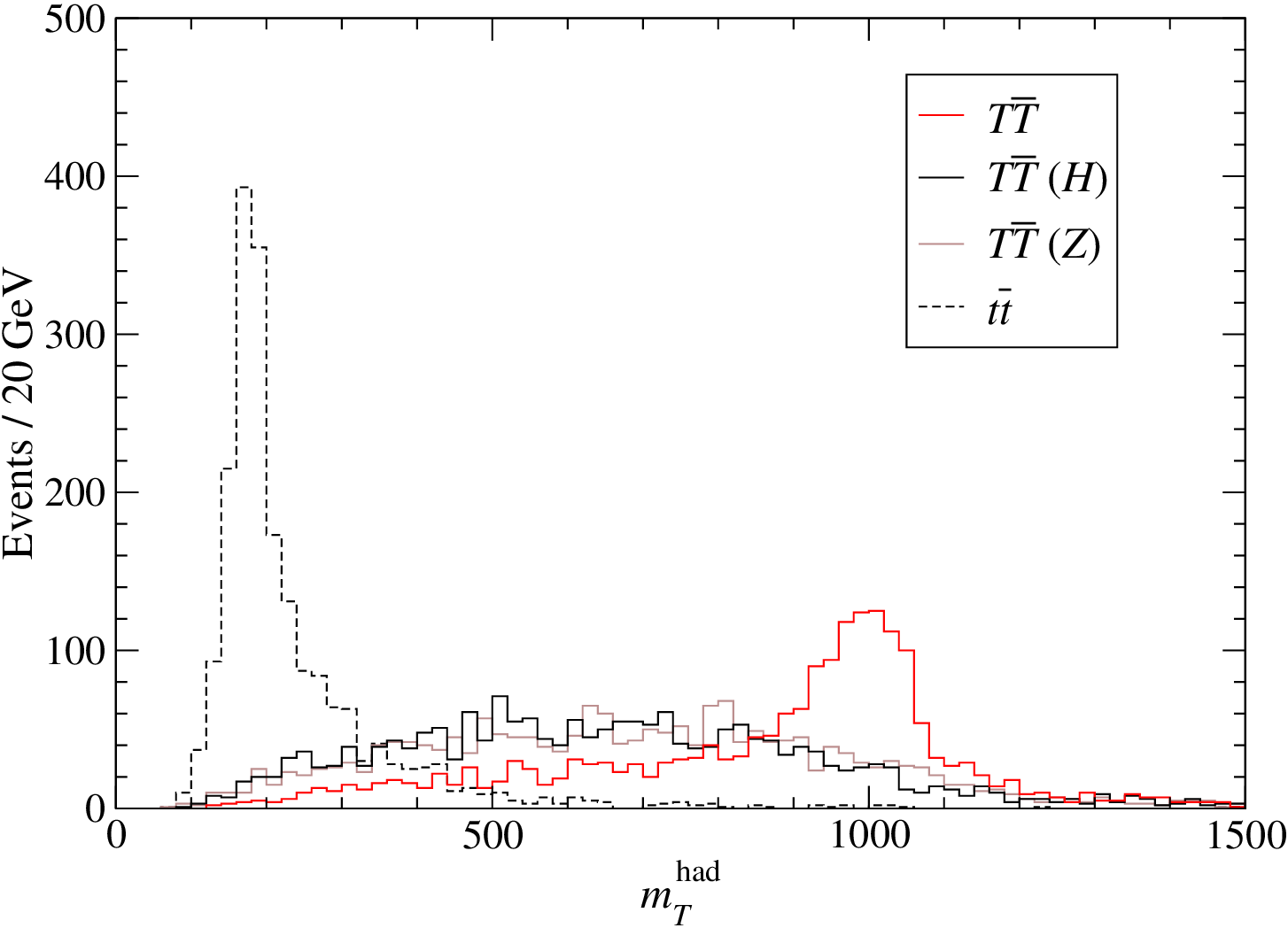,height=5.0cm,clip=} &
\epsfig{file=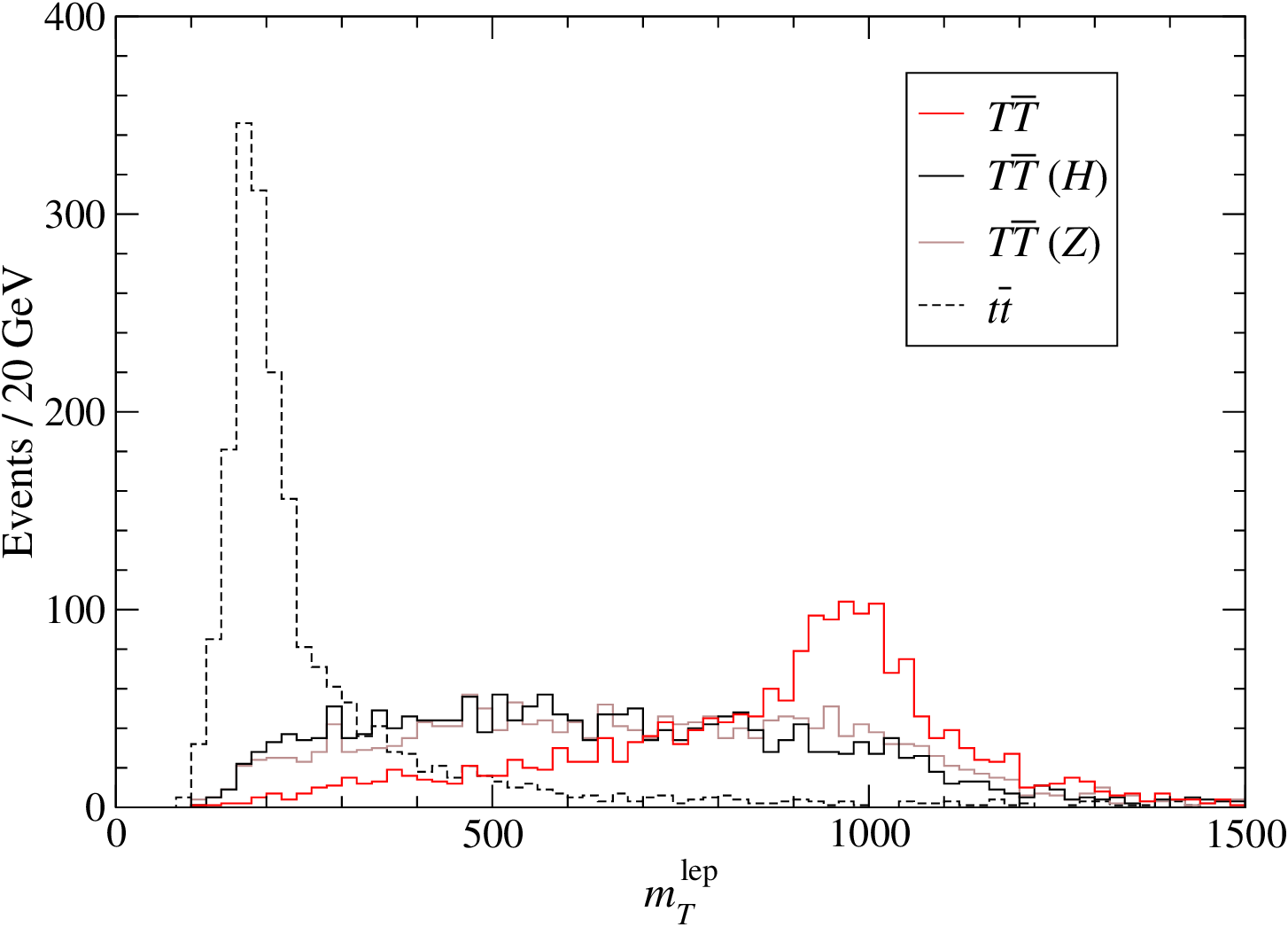,height=5.0cm,clip=} \\
(a) & (b) \\[0.5cm]
\multicolumn{2}{c}{\epsfig{file=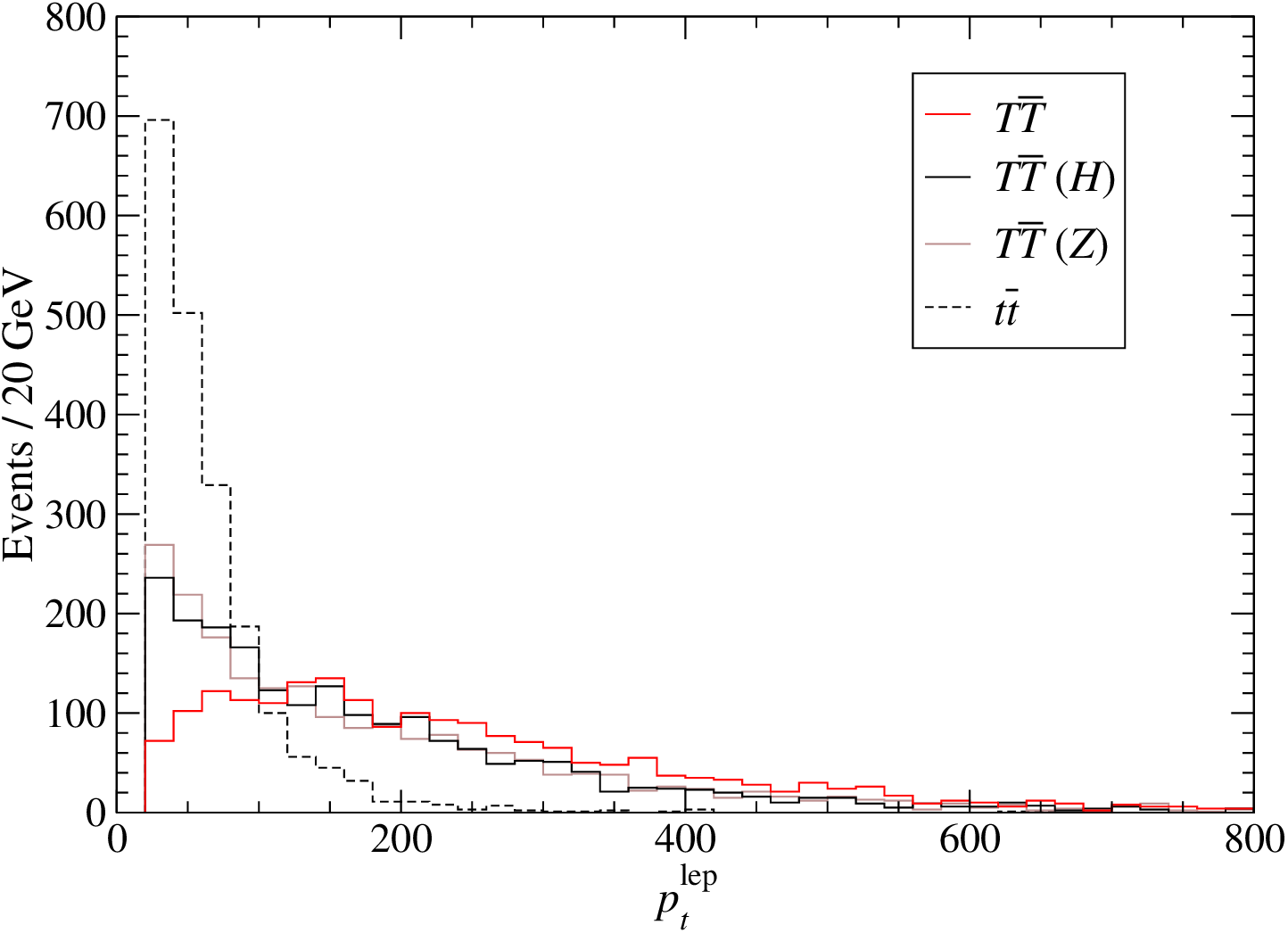,height=5.2cm,clip=}} \\
\multicolumn{2}{c}{(c)}
\end{tabular}
\caption{Reconstructed masses of the heavy quarks decaying hadronically (a) and
semileptonically (b), and transverse momentum of the charged lepton (c),
after pre-selection cuts.}
\label{fig:dist1000}
\end{center}
\end{figure}

\begin{figure}[p]
\begin{center}
\begin{tabular}{cc}
\epsfig{file=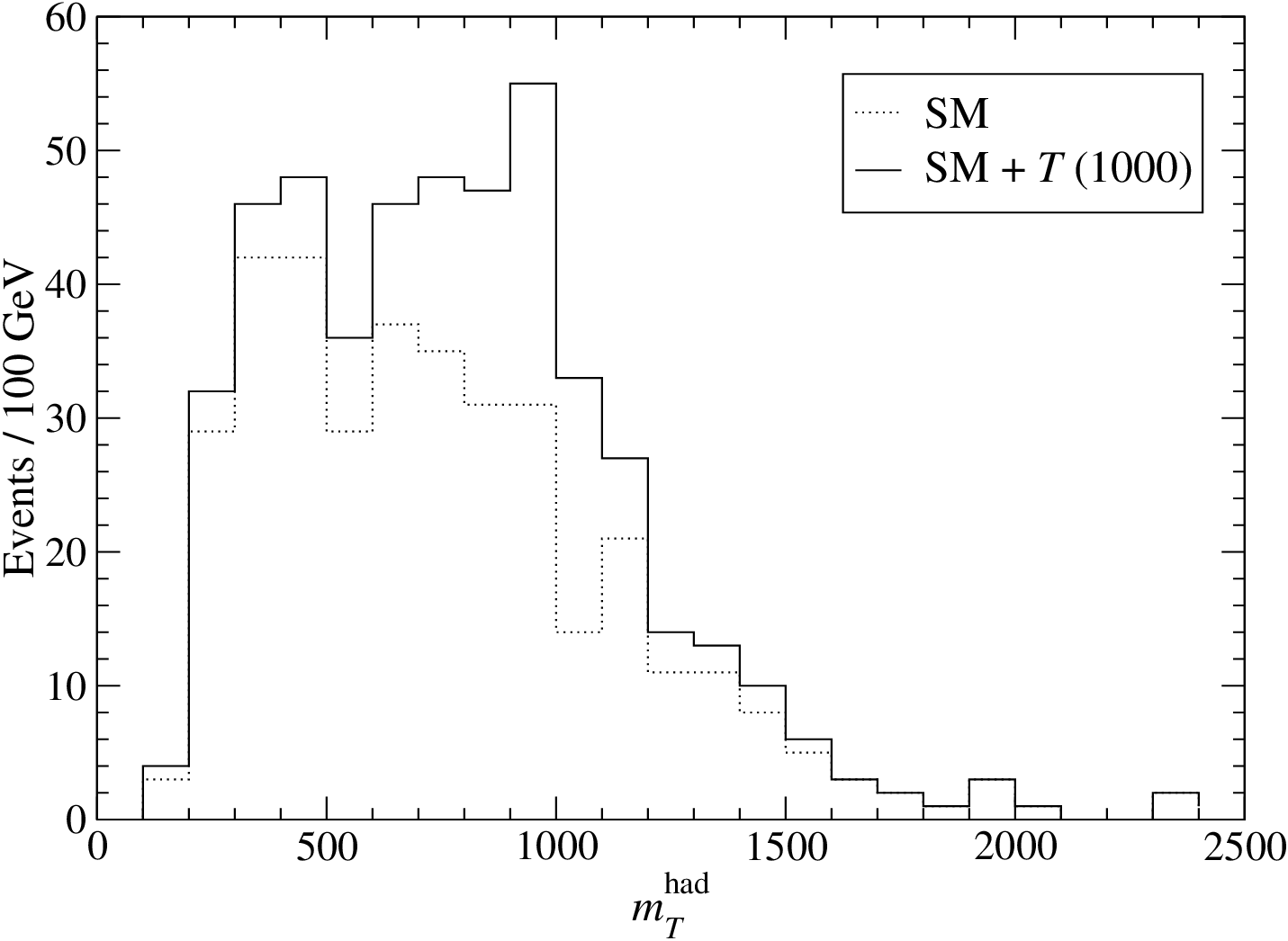,height=5.2cm,clip=} &
\epsfig{file=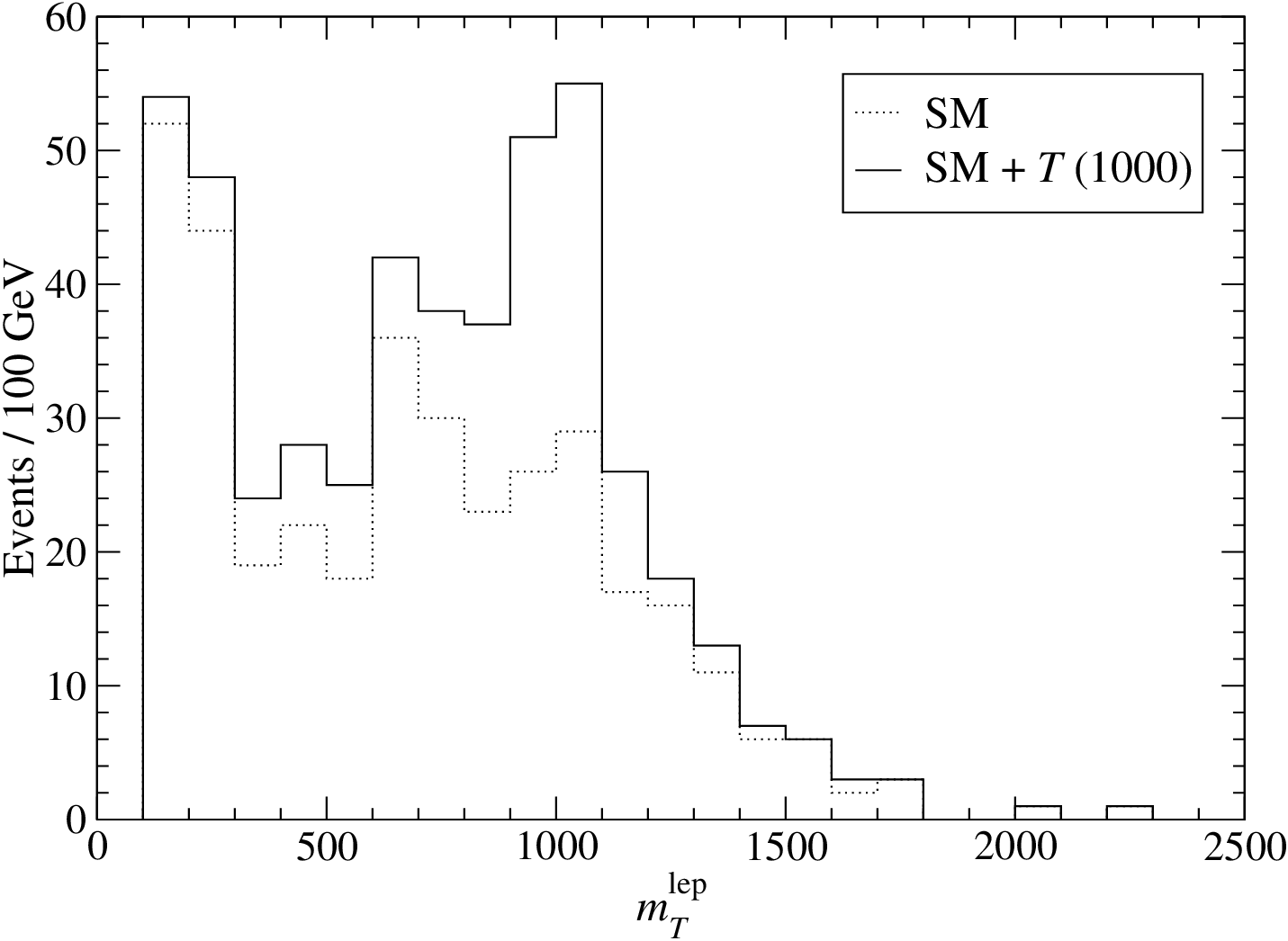,height=5.2cm,clip=} \\
(a) & (b)
\end{tabular}
\caption{Reconstructed masses of the heavy quarks decaying hadronically (a) and
semileptonically (b), after selection cuts. The
dashed lines correspond to the SM predictions, while the full lines represent
the SM plus a new 1 TeV quark.}
\label{fig:mT1000}
\end{center}
\end{figure}

The simulation in this case is done for 300 fb$^{-1}$ in the
high luminosity LHC run. The $b$ tagging efficiency assumed is of 50\%.
The reconstructed masses of the two quarks are shown in Fig.~\ref{fig:dist1000}
(a) and (b). For the $H$ and $Z$ signals the events are
spread over a wide range of $\mthad$ and $\mtlep$ values, and for this $T$ mass
a smaller fraction of events is in the vicinity of $m_T$.
The charged lepton transverse momentum is plotted in
Fig.~\ref{fig:dist1000} (c). For a heavy quark mass of 1 TeV, the lepton
momentum distribution from $T \to W^+ b \to \ell^+ \nu b$ has a very long tail.
Requiring $\ptlep \geq 200$ GeV eliminates 98\% of the $t \bar t$ background,
while keeping 51\% of the $W^+ b W^- \bar b$ signal. Our selection
criteria are in this case
\begin{align}
& \ptmax \geq 400 ~\mathrm{GeV} \,, \quad
\ptbmax \geq 300 ~\mathrm{GeV} \,, \nonumber \\
& \ptlep \geq 200 ~\mathrm{GeV} \,, \quad
50 ~\mathrm{GeV} \leq \ptmiss \leq 400 ~\mathrm{GeV} \,, \nonumber \\
& H_t \geq 1800  ~\mathrm{GeV} \,.
\label{ec:cut1000}
\end{align}
The mass distributions after these cuts are shown in Fig.~\ref{fig:mT1000}.
The number of events $N_\mathrm{cut}$ corresponding to each process can be read
in the first column of Table~\ref{tab:1000}, and the number of events in the
peak region
\begin{equation}
800 ~\mathrm{GeV} \leq \mthad \leq 1200 ~\mathrm{GeV} \;, \quad
800 ~\mathrm{GeV} \leq \mtlep \leq 1200 ~\mathrm{GeV}
\end{equation}
in the second column. The statistical significance of the $T$ signal is
$\mathcal{S}_0 = 9.09$, $\mathcal{S}_5 = 8.82$. The luminosity required to
discover a 1 TeV quark in pair production is 90 fb$^{-1}$.

\begin{table}[htb]
\begin{center}
\begin{tabular}{ccc}
Process & $N_\mathrm{cut}$ & $N_\mathrm{peak}$ \\
\hline
$T \bar T$ (1000) & 58.2 & 33.5 \\
$T \bar T$ (1000, $H$) & 39.6 & 7.8 \\
$T \bar T$ (1000, $Z$) & 21.0 & 5.1 \\
$t \bar t$ & 208 & 10 \\
$W b \bar b jj$ & 132 & 15 \\
$Z b \bar b jj$ & 19 & 1 \\
$t \bar b j$ & 3 & 0 \\
\end{tabular}
\caption{Number of events $N_\mathrm{cut}$ passing the selection criteria and
number of events $N_\mathrm{peak}$ passing the
selection cuts which are in the peak regions.}
\label{tab:1000}
\end{center}
\end{table}

\section{Discussion}
\label{sec:5}

We can compare the results obtained here with those from $T \bar T \to
W^+ b \, H \bar t,H t W^- \bar b$ decays in the final state $\ell \nu bbbb jj$,
requiring four $b$ tags \cite{higgs}. For $m_T = 500$ GeV and 10 fb$^{-1}$ of
luminosity, the latter channel has a similar sensitivity,
$\mathcal{S}_0 = 10.86$ , $\mathcal{S}_5 = 10.13$, which is achieved with
a good $b$ tagging performance, which suppresses the dangerous $t \bar t jj$
and $W +~$ jets backgrounds. We also note that the $\ell \nu bbbb jj$
channel has a slightly
larger branching ratio than $\ell \nu bb jj$ final states.
Combining both channels, a 500 GeV singlet can be discovered already with a
luminosity of 1.2 fb$^{-1}$.

\begin{figure}[htb]
\begin{center}
\epsfig{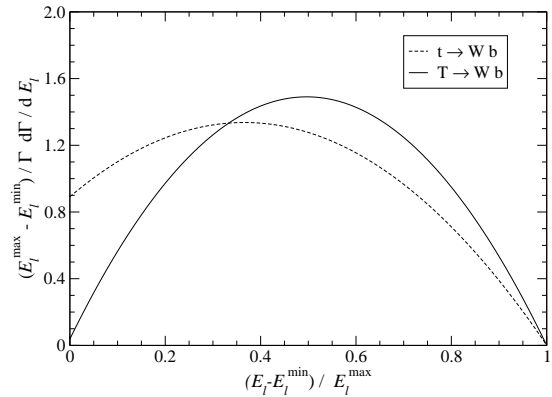}
\caption{Normalised energy distributions of the charged
lepton from $t$ and $T$ semileptonic decays.}
\label{fig:El}
\end{center}
\end{figure}

For $m_T = 1$ TeV the situation is rather different. The
$T \bar T \to W^+ b W^- \bar b$ signal is characterised by the presence of a
very energetic charged lepton, not only due to the higher energy available in
$T$ decays but also because of spin effects. Let us analyse these issues more in
detail.
For $m_T = 1$ TeV, the fraction of $W$ bosons with negative helicity
produced in $T$ decays is $\fm = 0.013$, compared to $\fm = 0.297$ for the top
quark, and then the charged lepton from $W$ decay is produced more towards the
$W$ flight direction. (The fraction of $W$ bosons with positive helicity $\fp$
is negligible in both cases due to angular momentum conservation; the zero
helicity fraction is then $\fz \simeq 1 - \fm$.)
This is reflected in the charged lepton energy distribution in the rest frame of
the decaying quark, which is \cite{asim}
\begin{eqnarray}
\frac{1}{\Gamma} \frac{d \Gamma}{d \el} & = & \frac{1}{(\emax -
\emin)^3} \left[ 3 (\el - \emin)^2 \, \fp + 3 (\emax - \el)^2 \, \fm
\right. \notag \\[1mm]
& & \left. + 6 (\emax - \el) (\el - \emin) \, \fz \right] \,,
\label{ec:dist2}
\end{eqnarray}
with $\emax$, $\emin$ the maximum and minimum charged lepton energies in the
rest frame of the decaying quark ($\emin = 18.5$
GeV, $\emax = 87.4$ GeV for $t$; $\emin = 3.2$ GeV, $\emax = 500$ GeV
for $T$). The normalised energy distributions of the charged lepton from $t$
and $T$ semileptonic decays are plotted in Fig.~\ref{fig:El}, where we have
scaled both axes to fully appreciate the spin effect. The different shape of the
curves is reflected in the $\ptlep$ distributions in Fig.~\ref{fig:dist1000}.

For $m_T = 1$ TeV, the requirement of high $\ptlep$ can be used to reduce
backgrounds significantly
in the $T \bar T \to W^+ b W^- \bar b$ mode but not in 
$T \bar T \to W^+ b \, H \bar t,H t W^- \bar b$ decays,
where half of the events have a charged lepton coming from a top (anti)quark.
(By the same reason, the $H$ and $Z$ contributions studied in this work are
suppressed.)
This might be partly compensated with a good $b$ tagging efficiency to reject
backgrounds, but at the high luminosity LHC run, where masses of the order of 1
TeV can be investigated, light jet and charm rejection is degraded. A
preliminary analysis has
shown that for $m_T = 1$ TeV and 300 fb$^{-1}$ of luminosity the significance
achieved in the $\ell \nu bbbbjj$ channel is $S/\sqrt B \lesssim 3$,
three times smaller than in the $\ell \nu bbjj$ mode analysed here.

\acknowledgments
I thank the organisers for an interesting and pleasant conference. 
This work has been supported by FCT through grant SFRH/BPD/12603/2003
and by a MEC Ramon y Cajal contract.

\end{document}